\documentclass[12pt]{article}
 
\def\Journal#1#2#3#4{{#1} {\bf #2}, #3 (#4)}

\def\PLB{{\em Phys. Lett.}  B}
\def\PRL{\em Phys. Rev. Lett.}
\def\PRD{{\em Phys. Rev.} D}
\def\ZPC{{\em Z. Phys.} C}
\def\APB{{\em Acta Phys. Polon.} B}
\def\EPC{{\em Eur. Phys. J.} C}
\def\JPG{{\em J. Phys.} G}
\def\PR{{\em Phys. Rep.}}

\newcommand{\lessim}{\raisebox{-0.8mm}%
{\hspace{1mm}$\stackrel{<}{\sim}$\hspace{1mm}}}
 
{\end{list}}
\newcounter{enumct}
\newenvironment{Enumerate}{\begin{list}{\arabic{enumct}.}%
{\usecounter{enumct}\setlength{\topsep}{0.2mm}%
\setlength{\partopsep}{0.2mm}\setlength{\itemsep}{0.2mm}%
\setlength{\parsep}{0.2mm}}}{\end{list}}
 
\newlength{\abstwidth}
\begin{document}
 
\sloppy
 
\pagestyle{empty}
 
\begin{flushright}
LU TP 99--23\\
hep-ph/9908408\\
August 1999
\end{flushright}
 
\vspace{\fill}
\begin{center}
{\LARGE\bf QCD Interconnection Effects\footnote{To appear in the
Proceedings of the International Workshop on Linear Colliders, 
Sitges (Barcelona), Spain, April 28 -- May 5, 1999}}\\[10mm]
{\Large Torbj\"orn Sj\"ostrand\footnote{torbjorn@thep.lu.se}} \\[3mm]
{\it Department of Theoretical Physics,}\\[1mm]
{\it Lund University, Lund, Sweden}\\[5mm]
{\large and} \\[5mm]
{\Large Valery A. Khoze\footnote{khoze@vxcern.cern.ch}}\\[3mm]
{\it INFN -- Laboratori Nazionali di Frascati,}\\[1mm]
{\it P.O. Box 13, I-00044 Frascati (Roma), Italy}\\[1mm]
\end{center}
 
\vspace{\fill}
 
\begin{center}
{\bf Abstract}\\[2ex]
\begin{minipage}{\abstwidth}
Heavy objects like the $W$, $Z$ and $t$ are short-lived compared with
typical hadronization times. When pairs of such particles are 
produced, the subsequent hadronic decay systems may therefore 
become interconnected. We study such potential effects at
Linear Collider energies.
\end{minipage}
\end{center}
 
\vspace{\fill}
 
\clearpage
\pagestyle{plain}
\setcounter{page}{1}

The widths of the $W$, $Z$ and $t$ are all of the order of 
2 GeV. A Standard Model Higgs with a mass above 200 GeV, as well 
as many supersymmetric and other Beyond the Standard Model particles
would also have widths in the \mbox{(multi-)GeV} range.  Not far from
threshold, the typical decay times 
$\tau = 1/\Gamma  \approx 0.1 \, {\mathrm{fm}} \ll  
\tau_{\mathrm{had}} \approx 1 \, \mathrm{fm}$.
Thus hadronic decay systems overlap, between pairs of resonances 
($W^+W^-$, $Z^0Z^0$, $t\bar{t}$, $Z^0H^0$, \ldots), so that the final 
state may not be just the sum of two independent decays. 
Pragmatically, one may distinguish three main eras for such
interconnection:
\begin{Enumerate}
\item Perturbative: this is suppressed for gluon energies 
$\omega > \Gamma$ by propagator/time\-scale effects; thus only
soft gluons may contribute appreciably.
\item Nonperturbative in the hadroformation process:
normally modelled by a colour rearrangement between the partons 
produced in the two resonance decays and in the subsequent parton
showers.
\item Nonperturbative in the purely hadronic phase: best exemplified 
by Bose--Einstein effects.
\end{Enumerate}
The above topics are deeply related to the unsolved problems of 
strong interactions: confinement dynamics, $1/N^2_{\mathrm{C}}$ 
effects, quantum mechanical interferences, etc. Thus they offer 
an opportunity to study the dynamics of unstable particles,
and new ways to probe confinement dynamics in space and 
time \cite{GPZ,ourrec}, {\em but} they also risk 
to limit or even spoil precision measurements \cite{ourrec}.

So far, studies have mainly been performed in the context of
$W$ mass measurements at LEP2. Perturbative effects are not likely
to give any significant contribution to the systematic error,
$\langle \delta m_W \rangle \lessim 5$~MeV \cite{ourrec}. 
Colour rearrangement is not understood from first principles,
but many models have been proposed to model 
effects \cite{ourrec,otherrec,HR},
and a conservative estimate gives 
$\langle \delta m_W \rangle \lessim 40$~MeV. 
For Bose--Einstein again there is a wide spread in models, and an 
even wider one in results, with about the same potential systematic
error as above \cite{ourBE,otherBE,HR}.
The total QCD interconnection error is thus below $m_{\pi}$ in 
absolute terms and 0.1\% in relative ones, a small number that 
becomes of interest only because we aim for high accuracy. 

More could be said if some experimental evidence existed, but
a problem is that also other manifestations of the interconnection 
phenomena are likely to be small in magnitude. For instance,
near threshold it is expected that colour rearrangement will deplete
the rate of low-momentum particle production \cite{lowmom}. Even with
full LEP2 statistics, we are only speaking of a few sigma effects,
however. Bose-Einstein appear more promising to diagnose, but so far 
experimental results are contradictory \cite{BEstatus}. 

One area where a linear collider could contribute would be by allowing 
a much increased statistics in the LEP2 energy region. A 100 fb$^{-1}$
$W^+W^-$ threshold scan would give a $\sim 6$ MeV accuracy on the $W$ 
mass \cite{Wilson}, with negligible interconnection uncertainty.
This would shift the emphasis from $m_W$ to the understanding of the 
physics of hadronic cross-talk. A high-statistics run, e.g. 50 fb$^{-1}$ 
at 175 GeV, would give a comfortable signal for the low-momentum 
depletion mentioned above, and also allow a set of other 
tests \cite{othertest,lowmom}. Above the $Z^0Z^0$ 
threshold, the single-$Z^0$ data will provide a unique $Z^0Z^0$ 
no-reconnection reference. 

Thus, high-luminosity, LEP2-energy LC (Linear Collider) runs would be 
excellent to {\em establish} a signal. To explore the {\em character} 
of effects, however, a knowledge of the energy dependence could give 
further leverage.

In QED, the interconnection rate dampens with increasing energy
roughly like $(1 - \beta)^2$, with $\beta$ the velocity of each $W$
in the CM frame \cite{QED}. By contrast, the nonperturbative QCD 
models we studied show an interconnection rate dropping more like 
$(1 - \beta)$ over the LC energy region (with the possibility of
a steeper behaviour in the truly asymptotic region). If only the central
region of $W$ masses is studied, also the mass shift dampens 
significantly with energy. However, if also the wings of the mass 
distribution are included (a difficult experimental proposition, 
but possible in our toy studies), the average and width of the mass 
shift distribution do not die out. Thus, with increasing energy, 
the hadronic cross-talk occurs in fewer events, but the effect in 
these few is more dramatic.    

The depletion of particle production at low momenta, close to 
threshold, turns into an enhancement at higher energies \cite{lowmom}. 
However, in the inclusive $W^+W^-$ event sample, this and other signals 
appear too small for reliable detection. One may instead turn to exclusive 
signals, such as events with many particles at low momenta, or at 
central rapidities, or at large angles with respect to the event axis. 
Unfortunately, even after such a cut, fluctuations in no-reconnection 
events as well as ordinary QCD four-jet events (mainly $q\bar{q}gg$) 
give event rates that overwhelm the expected signal. It could still 
be possible to observe an excess, but not to identify reconnections 
on an event-by-event basis. The possibility of some clever combination
of several signals still remains open, however.

Since the $Z^0$ mass and properties are well-known, $Z^0Z^0$ events 
provide an excellent hunting ground for interconnection. Relative to
$W^+W^-$ events, the set of production Feynman graphs and the
relative mixture of vector and axial couplings is different, however,
and this leads to non-negligible differences in angular distributions. 
Furthermore, the higher $Z^0$ mass means that a $Z^0$ is slower than
a $W^{\pm}$ at fixed energy, and the larger $Z^0$ width also brings
the decay vertices closer. Taken together, at 500 GeV, the reconnection 
rate in $Z^0Z^0$ hadronic events is likely to be about twice as large 
as in $W^+W^-$ events, while the cross section is lower by a factor of
six. Thus $Z^0Z^0$ events are interesting in their own right, but
comparisons with $W^+W^-$ events will be nontrivial.

As noted above, the Bose--Einstein interplay between the hadronic decay 
systems of a pair of heavy objects is at least as poorly understood as is 
colour reconnection, and less well studied for higher energies. In some 
models \cite{ourBE}, the theoretical mass shift increases with energy, 
when the separation of the $W$ decay vertices is not included. With this 
separation taken into account, the theoretical shift levels out at around 
200~MeV. How this maps onto experimental observables remains to be 
studied,  but experience from LEP2 energies indicates that the mass shift
is significantly reduced, and may even switch sign.

The $t\bar{t}$ system is different from the $W^+W^-$ and $Z^0Z^0$
ones in that the $t$ and $\bar{t}$ always are colour connected.
Thus, even when both tops decay semileptonically, 
$t \to b W^+ \to b \ell^+ \nu_{\ell}$, the system contains nontrivial 
interconnection effects. For instance, the total hadronic multiplicity, 
and especially the multiplicity at low momenta, depends on the opening
angle between the $b$ and $\bar{b}$ jets: the smaller the angle,
the lower the multiplicity \cite{topmult}. On the perturbative level, 
this can be understood as arising from a dominance of emission from 
the $b\bar{b}$ colour dipole at small gluon energies \cite{dipole}, 
on the nonperturbative one, as a consequence of the string 
effect \cite{string}.

Uncertainties in the modelling of these phenomena imply a systematic
error on the top mass of the order of 30~MeV already in the 
semileptonic top decays. When hadronic $W$ decays are included,
the possibilities of interconnection multiply. This kind of configurations
have not yet been studied, but realistically we may expect uncertainties
in the range around 100~MeV.

In summary, LEP2 may clarify the Bose--Einstein situation and
provide some hadronic cross-talk hints. A high-luminosity LEP2-energy 
LC run would be the best way to establish colour rearrangement, however.
Both colour rearrangement and BE effects (may) remain significant over 
the full LC energy range: while the fraction of the (appreciably) 
affected events goes down with energy, the effect per such event comes 
up. If the objective is to do electroweak precision tests, it appears 
feasible to reduce the $WW/ZZ$ ``interconnection noise'' to harmless 
levels at high energies, by simple proper cuts. It should also be 
possible, but not easy, to dig out a colour rearrangement signal at 
high energies, with some suitably optimized cuts that yet remain to 
be defined. The $Z^0Z^0$ events should display about twice as large 
interconnection effects as $W^+W^-$ ones, but cross sections are
reduced even more. The availability of a single-$Z^0$ calibration 
still makes $Z^0Z^0$ events of unique interest. While detailed studies 
remain to be carried out, it appears that the direct reconstruction of 
the top mass could be uncertain by maybe 100~MeV. Finally, in all of 
the studies so far, it has turned out to be very difficult to find a
clean handle that would help to distinguish between the different
models proposed, both in the reconnection and Bose--Einstein areas.
Much work thus remains for the future.

A copy of the transparencies of this talk, including all the figures
not shown here (for space reasons) may be found on\\ 
{\tt http://www.thep.lu.se/$\sim$torbjorn/talks/sitges99ww.ps}.\\
A longer writeup is in preparation.

\end{document}